# Design Features for the Social Web: The Architecture of *Deme*

Todd R. Davies and Mike D. Mintz

Symbolic Systems Program, Stanford University, Stanford, California, 94305-2150 USA
Davies@CSLI.Stanford.edu, MikeMintz@CS.Stanford.edu

**Abstract.** We characterize the "social Web" and argue for several features that are desirable for users of socially oriented web applications. We describe the architecture of *Deme*, a web content management system (WCMS) and extensible framework, and show how it implements these desired features. We then compare Deme on our desiderata with other web technologies: traditional HTML, previous open source WCMSs (illustrated by Drupal), commercial Web 2.0 applications, and open-source, object-oriented web application frameworks. The analysis suggests that a WCMS can be well suited to building social websites if it makes more of the features of object-oriented programming, such as polymorphism, and class inheritance, available to nonprogrammers in an accessible vocabulary.

## 1 Introduction

In early February of 2009, the social networking website Facebook.com changed its Terms of Service (ToS) agreement, apparently taking away users' right to remove, and thereby to revoke Facebook's license to, their content [16]. The move by Facebook caused considerable controversy, leading to a threatened lawsuit by privacy advocates, headline stories in major media, and an online revolt within Facebook by groups of its users. After several press communiqués and blog posts by Facebook representatives defending the new ToS failed to convince skeptics, the company changed course and reverted to its previous Terms of Service agreement [10].

The controversy over Facebook's ToS showed both the growing importance of the Web's social functions, as evidenced by the widespread media coverage it received, and also the widespread sensitivity of users to questions about who controls their data. This paper explores how the architecture of a web application can reflect these and other goals in the emerging, more socially oriented Web. After discussing use characteristics that distinguish social web applications, we define concepts and desired features for the social Web, and then describe the architecture of a new, socially oriented web content management system (WCMS) and extensible framework we call *Deme* (pronounced "deem"), which implements our approach to the challenges posed by our desiderata.

The unusual characteristics of the Web that pose engineering challenges were enumerated by Murugesan and Ginige [11]. There are various possible approaches to web engineering, including model-driven approaches [e.g. 5,15], principle-based approaches such as representational state transfer (REST)[8], and programming techniques such as agile programming [4]. Within specific programming language families, a notable development has been the widespread use of web application frameworks such as Ruby on Rails. A WCMS is a tool for building web applications without requiring website builders to have a background in computer programming.[1] Our approach combines the user experience goals of WCMSs with concepts and principles from object-oriented web application frameworks in an architecture built for the social Web.

## 2 Web 2.0 and the Social Web

The set of trends often referred to as "Web 2.0" – the second generation of web design and development – comprise roughly the following [cf 9,14]:

- websites that act as rich Internet applications (RIAs), with substantial and complex processing of HTTP requests;
- heavier use of client-side processing, i.e. more code loaded into the browser and less need for full page requests;
- larger websites supporting many interacting users, such as Wikipedia, Youtube, and Facebook; and
- availability and use of new, higher-level programming tools such as web application frameworks and rich Internet application frameworks.

In contrast to Web 2.0, the earlier generation ("Web 1.0") was based on the browser/client requesting full HTML pages or media files from a server directory, thin clients, relatively little interactivity with users, and little server-side code. The transition between the two was gradual, with some features of Web 2.0 being present even on some early websites.

Web content management systems pre-dated the introduction of the term "Web 2.0" in 2004 [14], and their development has lagged behind the phenomena cited above. The most popular general-purpose WCMSs today (Joomla/Mambo, Drupal, and Plone) were all first released in 2001-'02, and since then site administrators have gravitated toward newer versions of these tools rather than to newer WCMSs.[2] The greater emphasis on large (usually commercial) sites available to end users, on one hand, and developer tools requiring programming skills, on the other, has made nonprogrammers who administer websites less represented in the growth of the Web than both end users and programmers. But, as we will see, the characteristics of WCMSs are especially appropriate for the social Web.

While "Web 2.0" refers to a generation of technologies and trends that emerged in the second decade of the Web's existence, the term "social Web" refers to characteristics of

[1] See http://www.contentmanager.eu.com/history.htm.

[2] See http://mameou.wordpress.com/2008/05/31/dries-buytaert-vision-of-drupal/. Wordpress, which is more specific to blog management, was introduced a few years later and also been very successful. It could also be classified as a CMS, but is less general purpose than the other three tools cited.

web applications that pre-dated and will outlast the Web 2.0 era. The following use characteristics distinguish social web applications from non-social ones [2,12]:

**User-generated social content.** Social web applications enable site visitors to submit content that others can access, such as photos, their own profile data, links to other websites, and comments on other users' content.

**Social networking.** Users of social web applications join together in online groups and relationships (e.g. friends), which allow them to see identity-related information about the people to whom they are connected.

**Collaboration**. Users engage in conversations, co-creation of content (e.g. on wikis), collaborative filtering, and collective action.

**Cross-platform data sharing.** Increasingly, sharing content requires that a user be able to transfer data across sites, implying that the site on which the remote content is to be shared can interface correctly with the other site's data. When the remote data need to be processed locally, the two sites must agree on its meaning, which is a defining characteristic of the *semantic Web* [1,12].

A website need not exhibit all of the above characteristics in order to be considered social. For example, a newspaper blog may enable users to make comments, but with no support for networking through member profiles or collaboration. For our purposes, the main thing that defines the social Web is that it enables visitors, not just site administrators, to contribute some form of content that other users can access.

## 3 Content Management Concepts

We aim to show that our approach to content management provides advantages for the social Web over other approaches. To do this, we will define a set of dimensions through which content technologies can be compared. Consider the following concepts, with examples drawn primarily from traditional web concepts :

**Unit.** A chunk of content that can be referenced independently of other chunks, e.g. an HTML file/webpage.

**Subsegment.** A content segment or portion of a unit; for example, a semantic element in an HTML file.

**Unit type.** The classification of a unit of content that defines its subsegment structure and constrains what viewer code can be executed in order to display it, e.g. a MIME type.

**Behaviors.** A set of actions available to a user with respect to a given unit of content, e.g. the `GET`, `POST`, `PUT`, and `DELETE` methods of HTTP.

**Container.** A data structure for grouping multiple units together, e.g. a directory on a file server.

**Type structure.** The relationships between unit types, e.g. XHTML is a subtype of XML.

**Type-viewer matching.** A system for specifying which view code to invoke for a given unit type, e.g. the `preferences-applications` table in the Firefox browser.

**Relation specifier.** A way to represent and display a relationship between units, e.g. a hyperlink.

**Access control.** The system for specifying a user's abilities to perform actions, such as read and write, on a unit of content, e.g. through file system permissions.

**Addressing.** The means of specifying a particular unit, e.g. the URL of a page.

**Versioning.** A system for storing previous versions of a unit, e.g. an archive of old files.

**Deletion method(s).** A behavior of a unit that results in it becoming hidden or removed from use, e.g. the `rm` command in Unix and the `DELETE` method of HTTP.

**Software license.** A legal agreement with users specifying the rights and responsibilities of both the user and the provider of a site's software.

In the next section, we will argue for specific features along these dimensions that are most compatible with the social Web.

## 4 Guiding Features for Social Web Content Management

The use characteristics of the social Web suggest preferred features from the perspective of users (i.e. what users themselves are likely to want). These guiding features can be defined with respect to the content management concepts defined in section 3.

*Units should be* ***page independent.*** A web page may contain many pieces of data to which a user might want independent access. For example, the comments at the bottom of a blog posting should, logically, each be addressable and includable individually, as should any subset of them. Under HTML, by contrast, the page is the unit of reference. References to individual elements are possible through anchors or the DOM, but these are still tied to a particular page. Database tables are page-independent, but users do not generally have page-independent access to them. WCMSs generally allow users or administrators to define page independent units, such as the "nodes" of Drupal,[3] but commercial Web 2.0 sites such as Youtube do not permit end users this capability for each content unit, e.g. a particular comment. They require a user to refer to a unit through a URL.

*Subsegments should be* ***fully pointable***. Units themselves can be made up of parts or subsegments. But a user or content manager in a social web application may want a subsegment to be a reference to another unit, or to a subsegment within a unit. We call this feature "full pointability". The "creator" subsegment of a content unit might point, for example, to a unit representing a particular person. If the unit describing a person is divided into their name, email address, telephone number, etc., one of their friends might want to place their email address, without any of the other information about that person, in a list that will dynamically update whenever the email address changes. Although database-driven web applications generally allow references to other units, and the WCMS Drupal supports pointing to subsegments (node fields) in its Content Constructor Kit module,[4] in general commercial Web 2.0 sites do not allow the user to refer directly to content fields or subsegments.

*Unit types should be* ***polymorphic***. Polymorphism refers to the ability of a unit of data of one type to be treated as having a different type [3]. This is important for the social Web because, as noted in section 2, when one person shares content with another person, they may not be on the same platform, and so the code necessary to view a specific type may not be available to every user. Polymorphism exists for a unit when its type is a sub-type of an understood supertype (e.g. an HTML file with microformat markup can be rendered by any HTML interpreter). The Web 2.0 emphasis on both server- and client-side processing can break polymorphism for social web applications, because a given user's browser may not be compatible with a specific type of content. Cross-platform data sharing may also be impossible between different applications or even different installations of the same application.

[3] See http://drupal.org/node/19828.

[4] See http://drupal.org/project/fieldreference.

*Behaviors should be* ***extensible***. In general, a designer of a website will not anticipate all the possible actions and sequences of actions a user may want to do. For example, a user of a search engine may want to sort the results by a criterion for which there is no widget on the site, or a reader of a message board might want to view all the posts by people from a certain city. Extensibility implies that the user can create new behaviors for a given unit type. But this depends on an ability to modify either the data model or the view code. WCMSs such as Drupal provide this ability by allowing modifications of the open source code, and through optional contributed modules. Commercial Web 2.0 sites, by contrast, often leave the user without a way to add a desired behavior when the code is not available for modification. For example, users of many video sharing sites cannot sort videos by date.

*Containers should be* ***referential.*** In a tree-structured file system paradigm, the container (a folder or directory) stores a copy of each unit (a file), and every file must be stored in a folder. As Ted Nelson has pointed out [13], thinking this way results in unnecessary file duplication that can cause incompatibilities. Referential containers, by contrast, store only the addresses of content units, and units are stored separately. This form of container is better suited to the Web than value containers are, because different people have different ways of categorizing content, and with reference-based categories, they need not interfere with each other. Referential containers allow for a single point of storage, rather than copies that can be updated differentially and become inconsistent. Social web applications emphasize practices such as the sharing of tags or labels, which are reference containers.

*Type structure should be* ***inheritance hierarchical***. In Drupal, content types such as "article" and "event" are defined in a flat hierarchy, with configurable options, but without inheritance of structure from other types.[5] A site developer who uses Drupal expressed the need for type inheritance in the following blog comment in December 2008[6]:

> "…the structure of most of our content types is similar, or close enough that much of the template is the same for all of them. Most of the code in each template is repeated from one to the next. I really wish there was some kind of content type hierarchy or inheritance so those types of properties could be passed on to "children" content types."

The traditional structure of web pages is very nonhierarchical. Internet media types can associate different actions to different types of content in the browser, but they do not exist in an inheritance hierarchy. Content type inheritance, which is found in only a few enterprise CMSs such as Alfresco and Documentum (and not, to our knowledge, in general use Web CMSs) has been called by one blogger "the holy grail of content management".[7]

*Type-viewer matching should be* ***server-side specialized.*** The advantage of content typing is only fully realized if each content type is associated with view code that is tailored to that type. This makes it possible to tailor the user interface experience to the content type. WCMSs that allow content typing generally specialize the view by type to some extent, but when, as in Drupal, a user can create new content types by filling out a form, the view code must be generic enough across content types to allow the definition of a new type without writing code for an associated viewer. This limits the extent to which the viewer can be specialized. Again, web browsers traditionally render all content as pages of HTML, and Internet media/MIME types are handled differently by each browser. If a site administrator wants all (or nearly all) end users to be able to view content in a specialized way, the view code must be defined at the server level.

[5] See http://drupal.org/project/inherit regarding experimental content type inheritance in Drupal.
[6] See http://www.yelvington.com/node/517.
[7] See http://gadgetopia.com/post/6360.

*Relation specifiers should be* ***integrally unitizable.*** Relations between content units such as hyperlinks have traditionally been specified within one unit, pointing to another unit. This leads to a basic structure for links that is one-way, which can make it difficult to detect incoming links. Moreover, specifying relations within a unit usually requires that the user specifying the relation have write privileges for the target unit, and that these relations be visible to all users viewing the target unit, such as a web page. In social web applications, on the other hand, users may wish to specify relations between units in a way that is external to the related units. For example, a user may wish to insert a comment at a particular location in a document, which will have different associated permissions from those of the document itself. This is especially useful when referring to documents on other websites. Although even basic HTML supports linking to an external page, this kind of reference specifier is not integrated within the application, because it is not visible when viewing the referenced pages. A solution is to allow relations between specific locations in content units to be specified as units themselves, with their own permissions, and an integrated tool for displaying references when viewing an item. WCMSs such as Drupal generally support this[8] but not for the general case of relationships between locations within units,[9]

*Access control should be* ***fluid-granular***. Web applications generally provide much coarser control over who can view, edit, and delete content than does an operating system. But this type of control is what users typically want, because each piece of content is different. Moreover, the ideal privilege definitions are even more complex than in operating systems, since they can be defined for an arbitrary number of groups, with complicated rules of precedence, and for each subsegment (field) of each content unit. Commercial websites generally give limited control to the user to define these permissions, although social networking sites such as Facebook have evolved to be fairly granular.[10] Drupal embodies fluidity through the ability to define an indefinite number of roles, or packages of privileges, and its Content Creator Kit module makes field-level permissions available as well.[11] But the combination of fluidity (many distinctions between administrator and user) and granularity (control over each field of a unit) is very difficult to achieve and generally not found in commercial Web 2.0 sites.

*Addressing should be* ***domain independent***. As much as possible, content should be addressable independently of its path, so that links will not break if the content moves.

---

[8] See http://drupal.org/node/414018.

[9] A well-known advocate of more flexible reference specifiers, with support for two-way links and deep transclusion, has been Ted Nelson [13].

[10] See http://www.allfacebook.com/2009/04/facebook-privacy-limitations/.

[11] See http://drupal.org/node/310 and http://drupal.org/project/cck_field_perms.

This can be implemented through redirects, but that depends on the content owner's control over a domain, since URLs are tied to domain names. Commercial web applications generally do not support domain-independent addressing, but Drupal does support a limited version of it, internally to a site, through the node ID combined with the ability to move the database to another domain.

*Versioning should be **comprehensive.*** Since, on a social web site users are providing content, they may need access to earlier versions of a unit. This is built into Drupal[12] as well as wiki sites, but is generally not available on commercial Web 2.0 sites.

*Deletion methods should be **user controlled.*** In a social web application, a user uploads content to the host site in lieu of placing it on their own site. For commercial Web 2.0 sites, this means that the user's ability to remove content is limited by the tools provided to users, and by the ToS agreement. As in the Facebook ToS controversy, this can lead users to feel that they have lost control over their own data, and may pose privacy risks. A site should not unduly limit users' ability to delete their own data, e.g. by making true deletion impossible (as opposed to flagging the data as hidden in a database). Drupal makes true deletion available to users,[13] but commercial sites generally do not.

*Software licenses should be **free/open source.*** Another aspect of user control is the ability to inspect and modify the code. Although this generally requires moving data to one's own server space, and most users will not want to do it, a free/open source platform gives all users flexibility by enabling others to provide alternative hosting environments for their data.

## 5 The Deme Architecture

In this section, we describe the architecture of Deme,[14] our new WCMS and framework written in Django/Python, with a PostgreSQL database, licensed under the Affero GPLv3 license.[15] Recently, the term "content management *framework*" has been used, somewhat controversially, to denote "an application programming interface for creating a customized content management system".[16] We use the term "framework" to indicate that the system is designed to facilitate custom code development. Deme attempts to make available the concepts of object-oriented programming (OOP) to end users and nonprogrammer website administrators, using language that we believe will be more understandable to nonprogrammers. We define the terminology of Deme below with respect to concepts familiar to a technical audience. Desired features from section 4 are noted in ***bold italics***.

**Items and item types.** Units of content in Deme are stored in "items". An item is an instance of a particular "item type". The Deme item types are ***inheritance hierarchical***. If the `Person` item type inherits from the `Agent` item type, then any item that is a `Person` is also an `Agent`. Every item type ultimately inherits from the `Item` item type

---

[12] See http://drupal.org/node/70591.

[13] See http://agaric.com/note/disable-delete-regular-users.

[14] See http://deme.stanford.edu.

[15] See http://www.gnu.org/licenses/agpl-3.0.html.

[16] See for example http://en.wikipedia.org/w/index.php?title=List_of_content_management_frameworks&oldid=282731961

(which corresponds to the Object class in many programming languages). We allow multiple inheritance, and use it occasionally (e.g., `TextComment` inherits from both `Comment` and `TextDocument`). Deme items are stored in a database using object relational mapping (ORM)[17] with multi-table inheritance. For example, if our item type hierarchy is `Item` -> `Agent` -> `Person`, and our items are `Mike[Person]` and `Robot[Agent]`, then there will be one row in the `Person` table (for `Mike`), two rows in the `Agent` table (for `Mike` and `Robot`), and two rows in the `Item` table (for `Mike` and `Robot`). An abridged basic view of the Deme item type hierarchy is shown in Figure 1.

**Fig. 1.** The Deme item type hierarchy (abridged basic view).

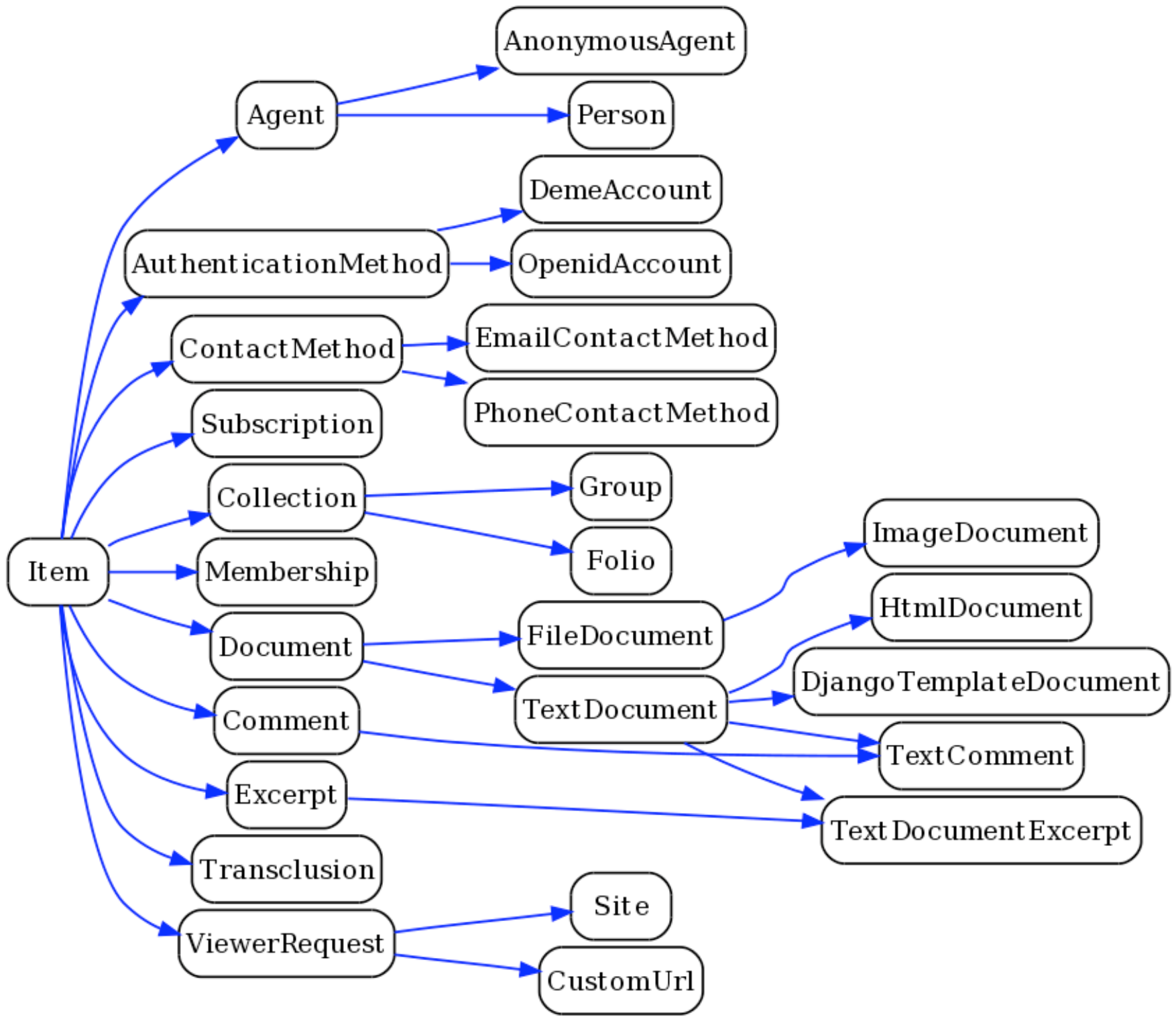

**Pieces.** Every item type defines the "pieces" (mapped to fields/columns in the database) relevant for that type's items, and item types inherit pieces from their supertypes. If `Item` defines the `description` piece, `Agent` defines no new pieces, and `Person` defines the `first_name` piece, then every person has a `description` and a `first_name`.

[17] See Scott W. Ambler's explanation ORM at http://www.agiledata.org/essays/mappingObjects.html.

**Piece types.** Every piece of an item has a type (e.g. `String`, `Integer`, and `Boolean`). Pieces can point to other items (foreign keys in the database), and can additionally specify which piece of another item is being pointed to (***fully pointable***). Pointing pieces are useful for defining relationships between items. For example, the `Item` item type has a `creator` piece pointing to the `Agent` that created it. Multiple items can point to a common item. Pieces cannot store data structures like lists. So rather than storing different contact methods as pieces of each `Agent`, we make `ContactMethod` an item type, and give it an `agent_pointer` piece. The contact methods for agent `123` are represented by all of the `ContactMethods` that have `agent_pointer` equal to `123`.

**Item IDs.** The most important piece of an item is the `id` (a ***page-independent*** primary key). Every item has a unique, immutable `id`. Items share the same `id` with their supertype versions (so `Mike`'s row in the `Person` table has the same `id` as `Mike`'s row in the `Agent` and `Item` tables). Pointing pieces are references to the `id` of the pointed-to item. Although not implemented yet, we plan to make available (optionally) "universal item `id`'s" through a reserved namespace approach like URNs or i-names, for ***domain-independent*** addressability across installations of Deme. Relations between items and pieces are shown for a portion of the item type hierarchy in Figure 2.

Other highlights of the Deme architecture include the following.

**Versioning.** For every item type, there is a ***comprehensive*** "old versions" table.

**Deleting items.** There are two ways to delete items: `deactivate` and `destroy`. Decativating is recoverable (through `reactivate`), but `destroy` is not (***user control***). The user interface ensures that deactivating happens before destroying.

**Collection.** An item type that represents an unordered ***referential*** set of other items, `Collections` use pointers from `Memberships` (which are items in their own right; ***integral unitizability***) to represent their contents, so multiple `Collections` can point to the same contained items. `Collections` "directly" contain items via `Memberships`, but they also "indirectly" contain items via chained `Memberships`.

**Transclusion.** An embedded reference from a location in a specific version of a `TextDocument` to another Item. A `Transclusion` [13] is a separate item (***integral unitizability***).

**Comment.** A unit of discussion about an item. Each `Comment` specifies the commented `item` and `item_version_number`. `Comments` can be associated with specific locations in a `TextDocument` via `Transclusions`.

**Excerpt.** An item that refers to a portion of another item or (in a planned future version) an external resource, such as a webpage (***full pointability***).

**Permissions.** Permissions define what actions an arbitrary group of `Agents` (***fluidity***) can and cannot do to each item and its pieces (***granularity***).

**Viewer types.** Deme takes advantage of the model-view-template architecture of Django. A viewer is a Python class that processes browser or API requests. Each viewer defines the item type it can accept (***server side specialization***), and multiple viewers can accept the same item type. Viewers that accept an item type will also accept subtypes of that item (***polymorphism***). Each viewer type defines a set of actions, e.g. `item_show`. Custom viewers and item types can define new actions (***extensibility***).[18]

---

[18] There are many other item types and architectural features not discussed above. For a full description, see http://deme.stanford.edu/static/docs/index.html.

**Fig. 2.** A detailed partial view of the Deme item type hierarchy. Solid connectors denote supertype-subtype inheritance. Dotted connectors denote pointers from pieces to items.

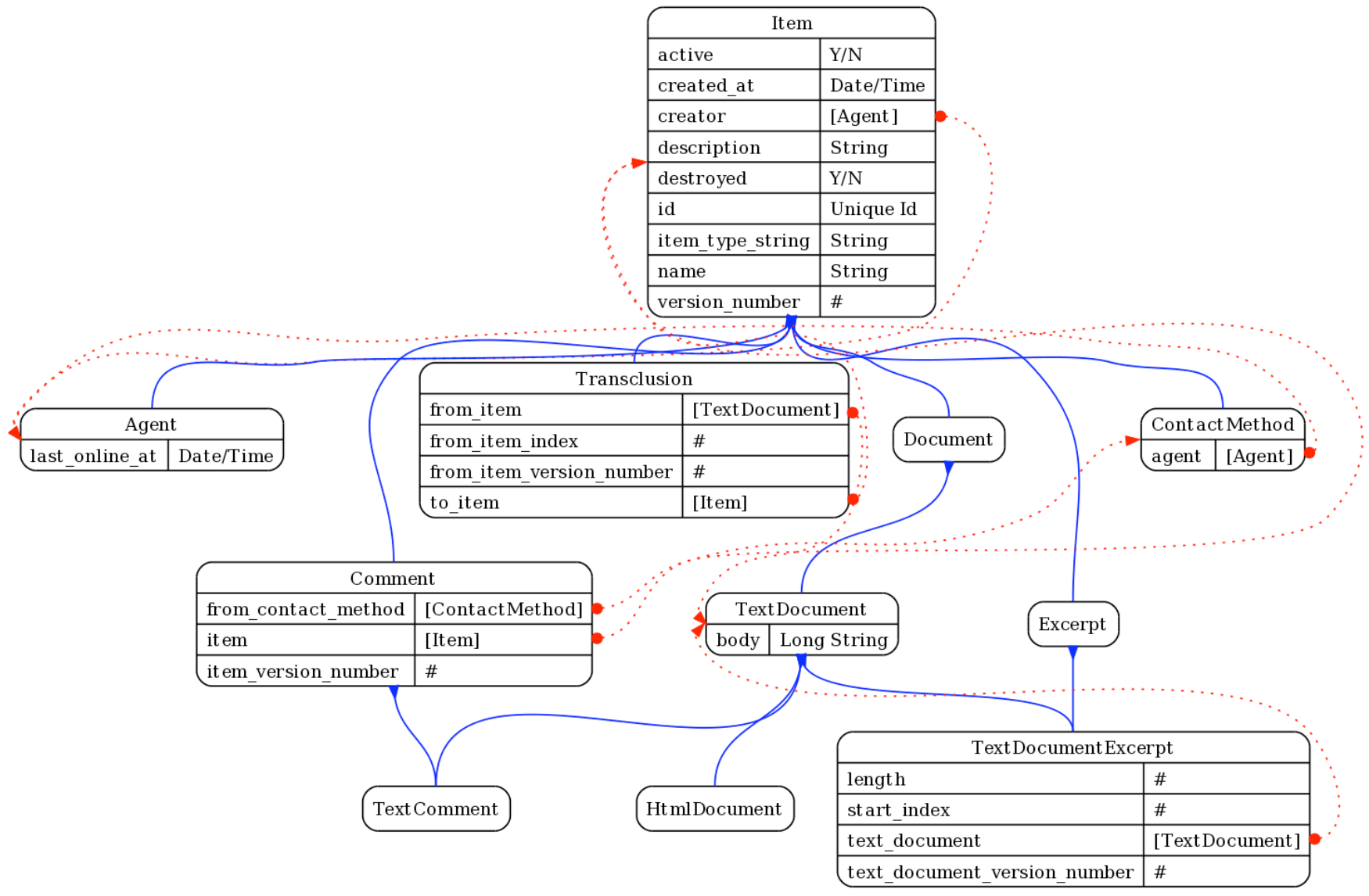

## 6 Comparing Deme With Other Technologies for the Social Web

Table 1 summarizes how Deme achieves the desired features for the social Web, by comparison with other web technologies: (a) file system-based "Web 1.0" sites (basic HTML); (b) the widely used WCMS Drupal; (c) commercial Web 2.0 sites such as Youtube, Facebook, and Myspace; and (d) object-oriented web application frameworks such as Ruby on Rails and Django.

On three dimensions (unit type, type structure, and addressing), the only other technology besides Deme that achieves the desired feature is OOP/web applicaton frameworks, which require programming skill. On eight dimensions (unit, subsegment, behaviors, relation specifiers, access control, versioning, deletion methods, and software license), all of the open-source approaches (Drupal, web frameworks, and Deme) achieve the desired features, but commercial Web 2.0 sites do not. The remaining two dimensions (container and type-viewer) are ones for which all of the technologies beyond basic HTML achieve the desired feature.

The social Web is especially associated with commercial Web 2.0 sites. But our analysis suggests that these sites do not meet users' needs as well as open-source technologies that give more control to users. Previous WCMSs, represented here by Drupal, generally exhibit more of the desired features than large commercial sites do, and they do not require a programming background to administer them. But they do not meet the desired social web criteria quite as well as OOP web frameworks do. The frameworks, on the other hand, require more programming skill. Deme makes available powerful OOP concepts from web frameworks to nonprogrammers for managing content, in a code base that is built for modification. We also believe that the terminology used in Deme will make it easier for nonprogrammers to learn than Drupal, but that remains to be tested empirically.

**Table 1.** Comparison of web technologies by content management concept. Approaches that make available the desired feature for each content concept are highlighted in **bold**.

| Content managment concept | Desired social feature | (a) File system/Web 1.0 HTML | (b) Web CMS (Drupal) | (c)Commercial Web 2.0 sites | (d) OOP /Web app frameworks | (e) Deme v0.9 WCMS |
|---|---|---|---|---|---|---|
| *unit* | ***page independent*** | file/page | **node** | photo, video, etc. | **object/row** | **item** |
| *subsegment* | ***fully pointable*** | semantic element | **field** | custom fields | **attribute/ field** | **piece, excerpt** |
| *unit type* | ***polymorphism*** | Internet media type | content type | custom types | **class** | **item type** |
| *behaviors* | ***extensible*** | HTTP methods | **menus** | widgets | **methods** | **actions** |
| *container* | ***referential*** | directory | **categories** | **tags/labels** | **container classes** | **collection** |
| *type structure* | ***inheritance hierarchy*** | MIME type /subtype | (flat) | (flat) | **class inheritance** | **item type hierarchy** |
| *type-viewer matching* | ***server-side specialized*** | browser application preferences | **views and modules** | **site-defined viewer** | **model-view separation** | **viewer types** |
| *relation specifiers* | ***integrally unitizable*** | one-way hyperlinks | **relation nodes** | limited bidirectional links | **relation objects** | **transclusions, memberships** |
| *access control* | ***fluid-granular*** | restricted directories | **admins and roles** | custom permissions | **customizable** | **permissions** |
| *addressing* | ***domain independent*** | URL | node ID | permalink | **object identiy** | **(universal) item id** |
| *versioning* | ***comprehensive*** | old files | **content versioning** | none or **wiki diffs** | **version control system** | **old versions table** |
| *deletion methods* | ***user controlled*** | file system delete | **node delete** | limited data removal | **file edit and delete** | **deactivate, destroy** |
| *software license* | ***free/open-source*** | default copyright | **GPLv2** | usually proprietary | **open source** | **Affero GPLv3** |

The version of Deme presented here is the latest step in a multi-year project aimed at creating a platform for deliberative interactions, e.g. document-centered discussion [6]. Future work will involve refining the interface to enable easier collaboration and commenting. The social Web is ultimately about fostering conversation. In the words of Cory Doctorow [7], "Conversation is king. Content is just something to talk about." Users are likely to continue to want this conversation to extend to an open dialogue about the social web platform itself. Even the most technically minded of tool providers should be pre-pared to justify their design and licensing choices to end users in relation to their needs and desires, and to provide technology that is responsive to user demands.

## Acknowledgments

We wish to thank Leo Perry, Ben Newman, Brendan O'Connor, Joseph Marrama, Jane Huang, and Ivan Sag for helpful contributions to this version of Deme.